\documentclass[aps,prb,twocolumn,superscriptaddress]{revtex4}
\usepackage{amsmath}
\usepackage{graphicx}
\begin{document}
\title{Finite Temperature Dynamical Correlations using\\
the Microcanonical Ensemble and the Lanczos Algorithm}
\author{M.W. Long}
\affiliation{School of Physics, Birmingham University, Edgbaston, 
Birmingham, B15 2TT, England}
\author{P. Prelov\v sek}
\author{S. El Shawish}
\affiliation{Faculty of Mathematics and Physics, University of Ljubljana,\\
and J. Stefan Institute, 1000 Ljubljana, Slovenia}
\author{J. Karadamoglou}
\author{X. Zotos}
\affiliation{Institut Romand de Recherche Num\'erique en Physique des 
Mat\'eriaux (IRRMA),\\
EPFL, 1015 Lausanne, Switzerland}
\date{\today}

\begin{abstract}
We show how to generalise the zero temperature Lanczos method for 
calculating dynamical correlation functions to finite temperatures. 
The key is the microcanonical ensemble, which allows us to
replace the involved canonical ensemble with a single appropriately 
chosen state; in the thermodynamic limit it provides the same physics 
as the canonical ensemble but with the evaluation of a single expectation 
value. We can employ the same system sizes as for zero
temperature, but whereas the statistical fluctuations present in small 
systems are prohibitive, the spectra of the largest system sizes are 
surprisingly smooth. 
We investigate, as a test case, the spin conductivity of the spin-1/2 
anisotropic Heisenberg model and in particular we present a comparison 
of spectra obtained by the canonical and microcanonical ensemble methods.
\end{abstract}


\maketitle

\section{Introduction}
The study of lattice quantum many body systems by the exact diagonalisation 
technique has proven popular at zero temperature ($T=0$) where only the 
ground state is required, but it is of less use at finite temperature. The
reason can be attributed to the different system sizes applicable, where for
spin-1/2 the ground state can be found for up to {\it N}$\sim $30 
lattice sites, but the entire spectrum can readily be achieved for systems 
only up to {\it N}$\sim $16. 
The intrinsic difficulties associated with applying the finite-size scaling 
method on such small 
systems severely limit finite temperature applications. At $T=0$
the continued fraction technique \cite{cfe1,cfe2} 
allows accurate calculations
of dynamical correlations using only the machinery of the Lanczos algorithm,
but unfortunately this technique has not been extended to finite $T$
where mostly full diagonalisation has been employed. As well as direct
applications of the canonical ensemble, there is also a 
hybrid method which employs the canonical representation of dynamical
correlation functions but uses a Lanczos basis to provide a set of
orthogonal states \cite{ftlm}. 
This method allows access to larger systems than are
accessible to full diagonalisation techniques but to smaller systems than the
current proposal, which does not need details of all the states even in the
Lanczos basis.  In this article we extend the $T=0$ formalism to
finite temperature by applying a {\it microcanonical ensemble} 
approach combined 
with the Lanczos method (MCLM) that provides smooth predictions 
for dynamical correlation functions at least at high temperatures.

The physical advance is to appreciate that in the thermodynamic limit the
microcanonical ensemble is equivalent to the canonical one \cite{ll,leb}, 
but for finite systems this is much easier to work with. The statistical
fluctuations engendered by the microcanonical choice are a drawback for
small systems but become controllable for large systems. In practice, as 
the finite $T$ calculations are much smoother, it is more natural to
contemplate applying finite-size scaling than for their $T=0$ counterparts.
The exponentially dense nature of a many particle spectrum in the bulk
is the property that smoothes our calculations, a characteristic that is lost
near the ground state where the spectrum is sparse.

Besides the computational interest of this proposal it is worth pointing 
out that, to our knowledge, no studies of the fundamental equivalence between 
the microcanonical and canonical ensemble for quantum dynamic correlations 
exist in the literature. Thus this work is a step in numerically 
exploring this basic postulate of nonequilibrium statistical mechanics;  
clearly, analytical studies are needed to clarify, for instance, the 
meaning of the microcanonical ensemble for a quantum system with dense 
spectrum as an average over a single quantum state (or a narrow window of 
states) and the finite size corrections inherent in this ensemble.

\section{Thermodynamic ensembles}

In this section we will discuss how an arbitrary probability distribution 
can be used, under reasonable assumptions, to represent the canonical 
ensemble in the thermodynamic limit.
The choice of a (unnormalized) distribution $\rho(\epsilon)$ in the 
thermodynamic limit 
can be examined by considering its Laplace transform, 

\begin{eqnarray}
e^{f(\tau)}=\int^{\infty}_0 d\epsilon~ e^{\epsilon \tau} \rho(\epsilon),
~~~\Re \tau < 0 \nonumber\\
\rho (\epsilon)=\int _{-i\infty-\eta}^{+i\infty -\eta}
\frac{d\tau }{2\pi i}
e^{-\tau \epsilon+f(\tau )},
\end{eqnarray}

\noindent
where $f(\tau)$ controls the properties of a distribution designated by
$\rho (\epsilon)$. As examples, the following choices of $f(\tau)$
lead to,

\begin{equation}
e^{f(\tau)}=\frac{1}{\beta - \tau} \mapsto 
\rho(\epsilon)=e^{-\beta\epsilon}
\end{equation}

\noindent
the canonical ensemble,

\begin{equation}
e^{f(\tau)}=e^{\lambda \tau} \mapsto 
\rho(\epsilon)=\delta(\lambda - \epsilon)
\end{equation}

\noindent
the microcanonical ensemble at energy $\lambda$,

\begin{equation}
e^{f(\tau)}=
e^{\lambda \tau+ \sigma^2(\tau -\beta)^2/2}
\mapsto 
\rho(\epsilon)=e^{-(\epsilon -\lambda)^2/2\sigma^2}
\end{equation}

\noindent
the ``Gaussian" ensemble at energy $\lambda$ and width $\sigma$.

If we examine the partition function $Z_{\rho}$, then,

\begin{equation}
Z_{\rho }=Tr \rho (\epsilon)=\int _{+i\infty }^{-i\infty }\frac{d\tau }
{2\pi i}e^{f(\tau )+\ln Z(\tau )}, 
\label{z}
\end{equation}

\noindent
where $Z(\tau )$ is the canonical partition function.  The physical idea
behind the thermodynamic limit is that the partition function becomes
immensely sharp when considered as a function of state space and becomes
dominated by the large number of states with the correct thermodynamics; in
practice fluctuating quantities can be replaced by their thermodynamic average
with negligible error. Mathematically, an integral such as (\ref{z}) may be
approximated in the asymptotic thermodynamic limit using the idea of `steepest
descents' with negligible error,

\begin{equation}
Z_{\rho }\propto e^{f(\beta ^*)} Z(\beta ^*).
\end{equation}

\noindent
Here $\beta ^*$ is chosen so that,
\begin{equation}
\frac{\partial f}{\partial \tau}(\beta ^*)=-\frac{1}{Z}
\frac{\partial Z}{\partial \tau}(\beta ^*)=\langle H\rangle 
\end{equation}

\noindent
and the average energy at the desired temperature is crucial.  For the
particular case of the microcanonical distribution 
$\lambda =\langle H\rangle $ 
and we need to employ states whose energy is the thermodynamic average as one
might naively guess.  Provided that $f(\tau)$ has only a weak dependence on
the system parameters, then the two partition functions are essentially
equivalent.  It is also clear that provided $f(\tau)$ has the required
properties that the `steepest descents' is a good approximation, then any
appropriate ensemble will provide the thermodynamic limit, for a particular
temperature.  It is quite natural to employ 
$f(\tau )=\lambda \tau +F(\tau)$ where $\lambda $ is extrinsic 
and $F(\tau )$ is intrinsic, in order to
limit towards the microcanonical ensemble.

\section{Dynamical correlations in the microcanonical and canonical ensemble}
The usually studied quantities of direct physical interest are the dynamic 
structure function,

\begin{equation}
S({\bf q},\omega )=\int _{-\infty }^{+\infty} dte^{i\omega t}
\langle X_{\bf q}(t) X_{-{\bf q}}(0)  \rangle,
\label{s}
\end{equation}

\noindent
and dynamic susceptibility,
\begin{eqnarray}
\chi({\bf q},\omega )&=&i\int _{0}^{+\infty} dte^{iz t}
\langle [  X_{\bf q}(t), X_{-{\bf q}}(0) ] \rangle,
\label{chi}
\end{eqnarray}

\noindent
where $z=\omega+i\eta$, the angle brackets denote a canonical ensemble 
thermal average and the 
commutator plays a central role in the linear response theory (or Kubo) 
formulation of transport.

The two quantities are related by the fluctuation-dissipation relation,

\begin{equation}
\chi''({\bf q},\omega)=\frac{1-e^{-\beta\omega}}{2}S({\bf q},\omega),
\label{fd}
\end{equation}

\noindent
where $\beta=1/k_BT$ is the inverse temperature. 
Note that $S({\bf q}, \omega)$ satisfies the symmetry relation 
$S (-{\bf q},-\omega )=e^{-\beta\omega}S ({\bf q},\omega )$ while the 
sum-rule,

\begin{eqnarray}
\frac{1}{2\pi}\int _{-\infty }^{+\infty} d\omega S ({\bf q},\omega )
=\langle  X_{\bf q} X_{-{\bf q}}\rangle,
\label{sr}
\end{eqnarray}

\noindent
makes it natural to consider the normalised to a unit area 
correlation function,

\begin{equation}
\hat S ({\bf q},\omega )\equiv \frac{S({\bf q},\omega )}{\langle
X_{\bf q} X_{-{\bf q}}\rangle }.
\label{norm}
\end{equation}

We have presented the dynamical correlation functions in the canonical
ensemble and now we will establish their form in the microcanonical one.
Starting from equation (\ref{s}) and employing solely the idea that
our distribution has a restricted energy $\lambda$ 
we can generate a correlation function 
$s({\bf q},\omega )$ in the microcanonical ensemble,

\begin{equation}
s({\bf q},\omega )=\int _{-\infty }^{+\infty} dte^{i\omega t}
\sum _m\langle X_{\bf q}\mid m\rangle \langle m\mid X_{-{\bf q}}\rangle 
e^{i(\lambda -\epsilon _m)t}.
\label{sm}
\end{equation}

\noindent
Here, we have used the relation, 

\begin{equation}
\langle  O U(H)\rangle \mapsto \langle O\rangle U(\lambda)
\end{equation}

\noindent
($U(H)$ a function of $H$) and a decomposition using the 
eigenbasis $\mid m\rangle$. 
The expression (\ref{sm}) integrates to provide,

\begin{equation}
s({\bf q},\omega )=2\pi\sum _m \langle X_{\bf q}\mid
m\rangle \langle m\mid X_{-{\bf q}}\rangle 
\delta (\omega +\lambda -\epsilon _m), 
\end{equation}

\noindent
that can be re-represented as the basic correlation in the microcanonical 
ensemble,

\begin{equation}
s ({\bf q},\omega )= - 2\lim _{\eta \mapsto 0}\Im 
\langle X_{\bf q}\left[ z -H+\lambda \right] ^{-1} X_{-{\bf q}}\rangle. 
\label{basic}
\end{equation}

\noindent
Notice that this expression is exact in the zero temperature limit 
where the expectation value is to be taken over the ground state wavefunction.

Now let us imagine that we could find a single eigenstate at will, with an 
energy arbitrarily close (in the thermodynamic limit) to a target energy, 
$\lambda$ say. It is in principle straightforward then to determine

\begin{equation}
s^*({\bf q},\omega )=-2\lim _{\eta \mapsto 0}\Im 
\langle *\mid X_{\bf q}\left[ z -H+\epsilon_* \right] ^{-1} 
X_{-{\bf q}}\mid * \rangle,
\end{equation}

\noindent
exactly as before, where $H\mid *\rangle =\epsilon _*\mid *\rangle $ is
the known eigenstate with $\epsilon _*\mapsto \lambda $. 
If the microcanonical 
ensemble is equivalent to the canonical ensemble and if a single eigenstate
is representative of the microcanonical one, then provided that
$\lambda =\langle H\rangle $ for the desired temperature, 
we can expect that

\begin{equation}
s^*({\bf q},\omega )\mapsto S ({\bf q},\omega )
\end{equation}

\noindent
in the thermodynamic limit. This amounts to the physical idea behind our
calculations.  

Furthermore, from (\ref{fd}), it follows that

\begin{equation}
\chi''({\bf q},\omega )=\frac{1-e^{-\beta\omega}}{2}s({\bf q},\omega)
\end{equation}

\noindent
and from the symmetry of $S({\bf q},\omega)$ in the canonical ensemble 
we can deduce that,

\begin{equation}
\ln \frac{s({\bf q},\omega )}{s(-{\bf q},-\omega )}
\mapsto \beta \omega;
\end{equation}

\noindent
this relation then provides an alternative, cross-checking technique 
for determining the temperature for a particular value of $\lambda $.
Although we might like to believe that a single eigenstate
corresponds to the microcanonical ensemble, based on a putative
ergodicity assumption for the eigenstate, in practice it is not possible 
to find such an eigenstate. So we relax the eigenstate hypothesis and go
back to a distribution of eigenstates close to the desired value $\lambda$.
We simply use the formalism as though we had such an eigenstate.

\section{The Lanczos method}

In principle we must construct a particular eigenstate with  
energy $\lambda$ that equals the canonical expectation 
value of the energy, $<H>=\lambda$, at the desired temperature. 
In practice, we employ the well known Lanczos algorithm that 
is an efficient way of diagonalising large Hamiltonians 
using as variational subspace (truncated basis) the set of states,

\begin{equation}
\big\{ \mid 0\rangle , H\mid 0\rangle ,...,H^{M_1}\mid 0\rangle \big\},
\end{equation}

\noindent
where $\mid 0\rangle $ is a (usually random) initial state and 
$M_1+1$ the number of Lanczos steps. 
To obtain an eigenstate close to energy $\lambda $ one might expect to use
the closest eigenstate to $\lambda $ in the truncated basis, but this is
totally incorrect. In practice, only the states at the edge of the
spectrum converge and the other `eigenstates' in the truncated subspace have
the suggested energies but are usually far from eigenstates. 

In order to apply the Lanczos method idea, one can simply push the energetic 
region of interest to the edge of the spectrum by choosing an appropriate 
new operator. One natural choice is to use,

\begin{equation}
K\equiv (H-\lambda )^2
\label{k}
\end{equation}

\noindent
which is positive definite and pushes the eigenstates with energy close to
$\lambda $ towards the minimal, zero, eigenvalue of $K$. 
Another way to understand this technique is
to consider expanding the ground state of $K$, that we will call 
$\mid\lambda \rangle$, as a probability distribution over
eigenstates. Choosing $\lambda$ establishes the appropriate mean for this
distribution but minimising $K$ corresponds to minimising the variance of the
distribution, and consequently localising the distribution near $\lambda $.

One can perform a Lanczos calculation based upon the operator 
$K$ or, more efficiently, one can evaluate the operator 
$K$ (now a pentadiagonal matrix) in a previously constructed Lanczos basis 
using $H$ (``L-projection" method). Note that,  

\begin{equation}
\langle (H-\langle H\rangle )^2\rangle =\langle (H-\lambda )^2\rangle
-\left( \langle H\rangle -\lambda \right) ^2\ge 0,
\end{equation}

\noindent
(the expectation value is over $\mid\lambda \rangle$) and so 
a small variance guarantees a narrow distribution of energies 
around $\lambda$.

In any Lanczos calculation the mathematical orthogonality between states
becomes lost at some stage as numerical errors build up.  In practice
only the well-separated converged states suffer from this disease and for us
these states, which are at the edge of the spectrum, do not gain any
significant weight in the correlation functions and so do not manifest in our
results.  The states at low frequency are all well behaved and maintain their
orthogonality.

It is straightforward to implement these ideas numerically, with a 
`double-Lanczos'
calculation; the first run through a Lanczos procedure of $M_1$ steps is 
employing the operator $K$ starting from a random state and
it is used to find the state $\mid\lambda\rangle$ which plays the 
role of the microcanonical distribution. 
The second run of $M_2$ steps through Lanczos is made using
$X_{\bf q}\mid\lambda\rangle$ as the initial state and then the resulting
tridiagonal matrix can be diagonalised to form the dynamical correlations
directly or by employing the continued fractions method which is numerically
more efficient but introduces a loss of resolution.

All the analysis so far has been subject to several caveats; firstly, that
the microcanonical ensemble is equivalent to the canonical one 
in the thermodynamic limit and in the context of quantum dynamic 
correlations. Secondly, that a single eigenstate is equivalent to the 
microcanonical ensemble and
thirdly that we can find such an eigenstate at will. 
The first two assumptions, as we have mentioned in the introduction, 
should be the focus of analytical studies as fundamental issues 
of nonequibrium statistical mechanics. 

Regarding the third assumption, it is clearly problematic as 
it is well known that although the Lanczos method 
converges quite easily at the sparse edges of the spectrum, in the denser 
inner regions of the spectrum, of interest at finite temperature, 
it takes the Lanczos procedure an
exponentially large number of iterations to converge.  A many-body spectrum has
an exponential number of states, e.g. for spin-1/2 the \# (States)$\sim 2^N$, 
and for a bounded Hamiltonian the eigenstates are compressed into an energy 
region that grows only linearly with system size. 
Although the low energy region maintains a
sparse density of states, the eigenstates become exponentially close
together in the area of interest and essentially become unattainable. 

At first sight this appears an insurmountable difficulty, but in practice 
this issue allows the technique its success. The first Lanczos procedure 
provides a single quantum state $\mid\lambda\rangle$, 
that is not an eigenstate, but which when decomposed in an 
eigenstate basis, it is represented by a narrow distribution 
$|a_n|^2$ around $\lambda$; 

\begin{eqnarray}
&&\mid \lambda \rangle=\sum_n a_n \mid n\rangle,
~~~ H\mid n \rangle =\epsilon_n \mid n \rangle
\nonumber\\
&&\langle \lambda \mid H \mid \lambda \rangle=\lambda,
\end{eqnarray}

\noindent
gives for the expectation value of an operator $O$,

\begin{eqnarray}
\langle \lambda \mid O \mid \lambda  \rangle&=&
\sum_n  \mid a_n \mid ^2 \langle n \mid O \mid n \rangle\nonumber\\
&+&\sum_{n\ne m} a_m^* a_n  \langle m \mid O \mid n \rangle.
\label{rpa}
\end{eqnarray}

This state, used in the evaluation of expectation values, acts as a 
statistical average over an energy window. It is important to notice, 
that by employing a single quantum state (not eigenstate) 
for evaluating an expectation value (as a substitute for a statistical average 
over a narrow energy window of eigenstates), we assume that the appearing 
off-diagonal terms (second term in eq.(\ref{rpa})) cancel each other.
This assumption can be justified (and numerically verified) by invoking 
a random phase decomposition of the used quantum state.

From this discussion we can expect two types of fluctuations in the 
obtained spectra; first, intrinsic fluctuations due to the finite size of 
the system, present even when a single eigenstate is used for the evaluation 
of the expectation value. Second, statistical fluctuations entering by  
the off-diagonal terms in eq. (\ref{rpa}) due to the use of a single 
pure state that is not an eigenstate; this type of fluctuations 
can be reduced by averaging over orthogonal states $\mid\lambda\rangle$ 
(e.g. corresponding to different translational symmetry $k-$ subspaces as 
we will show below).

\section{Convergence of projection}
In the following we present a test on the rate of convergence of the 
projection to a single quantum state with energy close to $\lambda$. 
Due to the innate complexity of an implicit scheme like Lanczos, we develop 
the theory of a simpler technique briefly to exhibit the ideas. 

A rather simple
method of numerically solving for the ground state is by an iterative
sequence of applications of the scaled Hamiltonian. For us this amounts to
iterative applications of the operator,

\begin{equation}
P=1-\left( \frac{H-\lambda }{\mu}\right) ^2,
\label{p}
\end{equation}

\noindent
where $\mu$ is chosen to be large enough so that $\mu ^2>
(\epsilon _n-\lambda)^2$ for the full spectrum. 
Repeated applications of this operator
exponentially suppresses all states except those for which $\epsilon _n\sim
\lambda $ which remain unaffected.  We can start out with a set of random
states and then for $M$ applications of our operator we can build a
distribution,

\begin{equation}
\rho _M(H)=\sum _\psi P^M\mid \psi \rangle
\theta \left[ P \right] \langle \psi \mid P^M
\end{equation}

\noindent
($\theta(P)$ is the step function) and if we were to perform an average over 
an orthogonal basis, $\mid \psi \rangle $, then this would converge to,

\begin{equation}
\rho _M(H)\mapsto \theta \left[P\right] P^{2M}.
\end{equation}

\noindent
Elementary analysis provides:
\begin{widetext}
\begin{equation}
\rho _M(H)=\int _{i\infty }^{-i\infty }\frac{d\beta }{2\pi i}
\exp\left[ \lambda \beta -\beta H\right] (2M)!2^{2M}2\mu 
\left[\frac{1}{x}\frac{d}{dx}\right]^{2M}
\frac{\sinh x}{x}\vert _{x=\beta \mu }.
\end{equation}
\end{widetext}

\noindent
In the limit that $M\mapsto \infty $ we find that,

\begin{equation}
f(\beta )\mapsto \lambda \beta +\frac{\mu ^2\beta ^2}{2(4M+3)}
+O\left(\frac{1}{M^2}\right)
\end{equation}

\noindent
and we converge to a narrow Gaussian probability distribution,

\begin{equation}
\rho _M(H)\sim \exp \left[ -\frac{(H-\lambda)^2}{2\mu ^2}(4M+3)
\right]. 
\end{equation}

\noindent
The width of this distribution is under our control,

\begin{equation}
\langle (H-\lambda)^2\rangle \sim \frac{\mu ^2}{4M+3}\sim 
\frac{W^2N^2}{4M+3},
\end{equation}

\noindent
where $W$ is the natural energy scale for the model and we see that $M$ needs
to scale with the square of the system size $N$ to maintain resolution.

\begin{figure}
\includegraphics[width=8.0 cm]{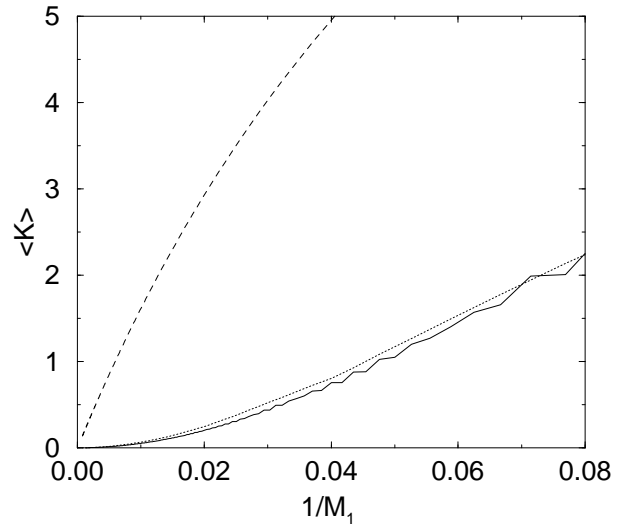}
\caption{Convergence properties of different Lanczos projection procedures: 
(i) dashed line, using eq.(\ref{p}), 
(ii) dotted line, using $K=(H-\lambda)^2$, 
(iii) continuous line, ``L-projection" (see text).}
\label{fig1}
\end{figure}

The Lanczos method is clearly much more sophisticated and provides a much
narrower distribution.  We have examined the distribution obtained in a
Lanczos calculation and we find that it is well represented by a Gaussian
distribution with a variance controlled by the `eigenvalue' of $K$ attained
by the calculation.  In practice this is about two orders of magnitude better
in energy than the result obtained from the projection analysis 
(eq.(\ref{p})), which however it is analytically controllable; indeed,
we find that the Lanczos method scales 
as $\langle K\rangle \propto {M_1}^{-2}$ so that 
the intrinsic resolution, $\sigma=\sqrt{\langle K \rangle}$,  
is inversely proportional to the number of
iterations. The convergence properties of the three schemes we discussed 
are depicted in Figure \ref{fig1} for a representative calculation of the 
study that we present in the next section.

The application of the technique should now be transparent; employing a single
random state, or averaging over a sequence of orthogonal random states, one
performs a first Lanczos calculation of $M_1$ steps to find the approximate 
ground state $\mid\lambda \rangle$ for
the operator $K=(H-\lambda )^2$. The value of $\lambda$ must be pre-selected
so that $\lambda =\langle H\rangle $ for the chosen temperature; 
several techniques are available for reliably determining this energy versus 
temperature relation as the Bethe Ansatz (for integrable 
systems), the finite temperature Lanczos (FTLM) \cite{ftlm}, the 
Transfer Matrix Renormalization Group (TMRG) or Quantum Monte Carlo method.
The degree of
convergence can be measured using the eventual `eigenvalue' of $K$;  
it plays the role of the variance of the chosen distribution
and its square-root is an intrinsic energy resolution $\sigma$. This
scale, $\sigma $, can never drop below the distance to the nearest eigenvalue. 
For a usual size system, e.g. $N>16$, and temperature, this limit is
unattainable but a resolution of $\sigma \sim 0.01$ 
($\langle K \rangle \sim 0.0001$) is readily
available with a thousand or so $M_1$ iterations. 

Once one has found this state $\mid \lambda \rangle$,  
that plays the role of the state 
$\mid * \rangle$, a second Lanczos projection sequence is generated 
employing the state $X_{\bf q} \mid \lambda \rangle$ as the initial state. 
The resolution of the eventual result is controlled by the intrinsic
dependence on the microcanonical ensemble and the degree of convergence
measured by $\sigma$. This can be seen from relation 
(\ref{basic}) as the eigenstates over which the state $\mid \lambda\rangle$ 
is decomposed 
have a spread in energy $\sigma$ with respect to the reference energy 
$\lambda$. The resolution also depends on the convergence achieved in the
2nd Lanczos procedure  where the number of iterations $M_2$ denotes the
finite number of poles which are used to try to represent the dynamical
correlations. At the sparse edges of the spectrum these poles denote the
eigenvalues of the system but in the bulk of the spectrum, when grouped into 
bins of a given frequency width, they are fairly uniformly spread and 
offer a further natural energy resolution for the calculation.  

More Lanczos steps provide more poles and a finer spectral `grid' for the
correlation functions, until the graininess of the real system is achieved.
We have elected to use a few thousand poles in our calculations with very
little improvement obtained by increasing this number as we 
shall see. The final resolution is self-imposed and is the $\eta$ of 
(\ref{basic}) which we choose to be of order of the spectral grid in order 
to smooth our calculations.

\section{Application on the spin-1/2 Heisenberg model}
We are now in a position to test our proposed technique and uncover its
strengths and weaknesses.  We have chosen to investigate the finite 
temperature dynamics of the prototype spin-1/2 Heisenberg model (equivalent to 
the fermionic ``$t-V$" model). This choice was
dictated by its central role in low dimensional quantum magnetism; 
an exact solution of the thermodynamics and 
elementary excitations is known using the Bethe Ansatz method \cite{korepin}, 
the spin dynamics probed by NMR is of current experimental interest 
\cite{takigawa,thurber} and several numerical and analytical studied have 
been devoted to the study of finite temperature dynamic correlations 
\cite{zp,mccoy,nma,z,gros}. The Hamiltonian is given by,

\begin{equation}
H=\sum_{l} h_l=J\sum_{l=1}^{N} (S_l^x S_{l+1}^x +
S_l^y S_{l+1}^y + \Delta S_l^z S_{l+1}^z),
\label{heis}
\end{equation}
where $S_l^{\alpha}~~(\alpha=x,y,z)$ are spin-1/2 operators on site $l$ 
and we take $J$ as the unit of energy and frequency ($\hbar=1$).

In particular, we will look at the high temperature spin conductivity in the 
antiferromagnetic regime, $J,~\Delta >0$, for which several studies exist 
and some exact results are known \cite{z}.
To discuss magnetic transport, we first define the relevant spin
current, $j^z$, by the continuity equation of the
corresponding local spin density $S^z_l$ (provided the total $S^z$
component is conserved),

\begin{equation}
S^z=\sum_l S_l^z,~~~~
\frac{\partial S_l^z}{\partial t}+\nabla j_l^z=0.
\label{contjs}
\end{equation}

\noindent
Thus, we obtain for the spin current $j^z$, 
(that plays the role of the operator $X_{\bf q}$), 
\begin{equation}
j^z=\sum_l j^z_l=J\sum_{l} (S_l^x S_{l+1}^y-S_l^y S_{l+1}^x).
\end{equation}

\noindent
The real part of the ``spin conductivity" $\sigma'(\omega)$ 
(corresponding to the charge conductivity of the fermionic model)  
includes two parts, the Drude weight $D$ and the 
regular part $\sigma_{reg}(\omega)$ \cite{zp,znp},

\begin{equation}
\sigma'(\omega)=2\pi D \delta(\omega)+\sigma_{reg}(\omega).
\label{real}
\end{equation}

\noindent
The regular contribution is given by,

\begin{equation}
\sigma_{reg}(\omega)=\frac{1-e^{-\beta\omega}}{\omega}\frac{\pi}{N}
\sum_{n\neq m} p_n |<n|j^z|m>|^2\delta(\omega-\omega_{mn})
\label{sigma}
\end{equation}

\noindent
where $p_n$ are the Boltzmann weights and 
$\omega_{mn}=\epsilon_m-\epsilon_n$. 

To compare the presented data on the 
conductivity we normalize them using the well known optical sum-rule that 
in the $\beta\rightarrow 0$ limit takes the form,

\begin{equation}
\int_{-\infty}^{+\infty} d\omega \sigma_{reg}(\omega)+2\pi D=
\beta\frac{\pi}{N}\langle {j^z}^2 \rangle.
\end{equation}

\noindent
The normalized conductivity, $\sigma(\omega)$, in this high temperature 
limit is given by,

\begin{equation}
\sigma(\omega)=
\frac{\sum_{n\neq m} |<n|j^z|m>|^2\delta(\omega-\omega_{mn})}
{\langle {j^z}^2 \rangle},
\end{equation}

\noindent 
that can be calculated using our microcanonical ensemble procedure by,

\begin{equation}
\sigma(\omega) \mapsto 
-\lim _{\eta \mapsto 0}\frac{\Im
 \langle \lambda\mid j^z \frac{1}{z -H+\lambda} j^z \mid \lambda \rangle }
{\pi \langle \lambda \mid {j^z}^2 \mid \lambda \rangle}.
\end{equation}

\noindent
In principle this expression includes also the contribution from the zero 
frequency Drude weight $\delta-$function, but in practice as the second 
Lanczos procedure cannot fully converge, the Drude peak appears as a low 
frequency contribution. As we will discuss below, sorting out this low 
frequency part, in general allows us to reliably extract the Drude 
weight value. 

\begin{figure}
\includegraphics[width=8.0 cm]{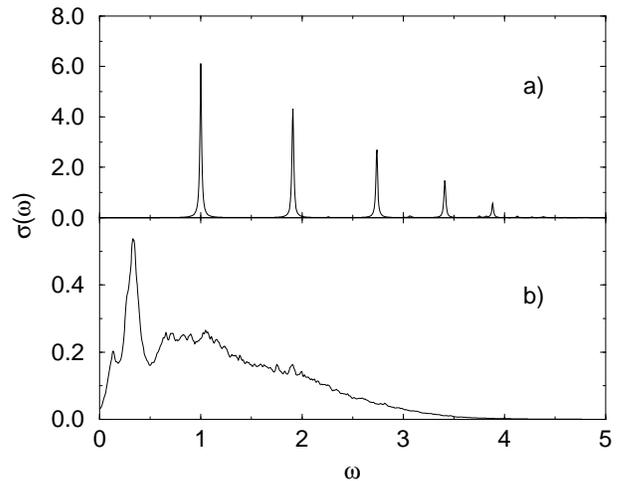}
\caption{
Microcanonical calculations for $N=26$, $\Delta =2$,
$\eta =0.02$; (a) $T=0$, (b) $\beta \rightarrow 0$.} 
\label{fig2}
\end{figure} 

In general, we can employ the translational 
symmetry of the Hamiltonian and study spectra in a given pseudomomentum 
$k-$ subspace or average the results over different $k-$ subspaces;
in the following we typically employ $M_1=1000$ and $M_2=4000$ Lanczos  
iterations at $\beta \rightarrow 0$ unless otherwise stated.
In Figure \ref{fig2} we compare a zero temperature \cite{cfe2} 
with an infinite temperature ($\beta\rightarrow 0$) 
calculation for a fairly large system in the $k=0$ subspace. 
The zero temperature calculation finds a few poles with exact weights 
whereas the infinite temperature
calculation provides a much smoother result.

There is clear
structure in the infinite temperature result but also apparently some noise.
To interpret this result we must consider the issue of the veracity 
of the microcanonical ensemble for such
small systems namely the extent to which the microcanonical ensemble is 
equivalent to the canonical one. 

\begin{figure}
\includegraphics[width=8.0 cm]{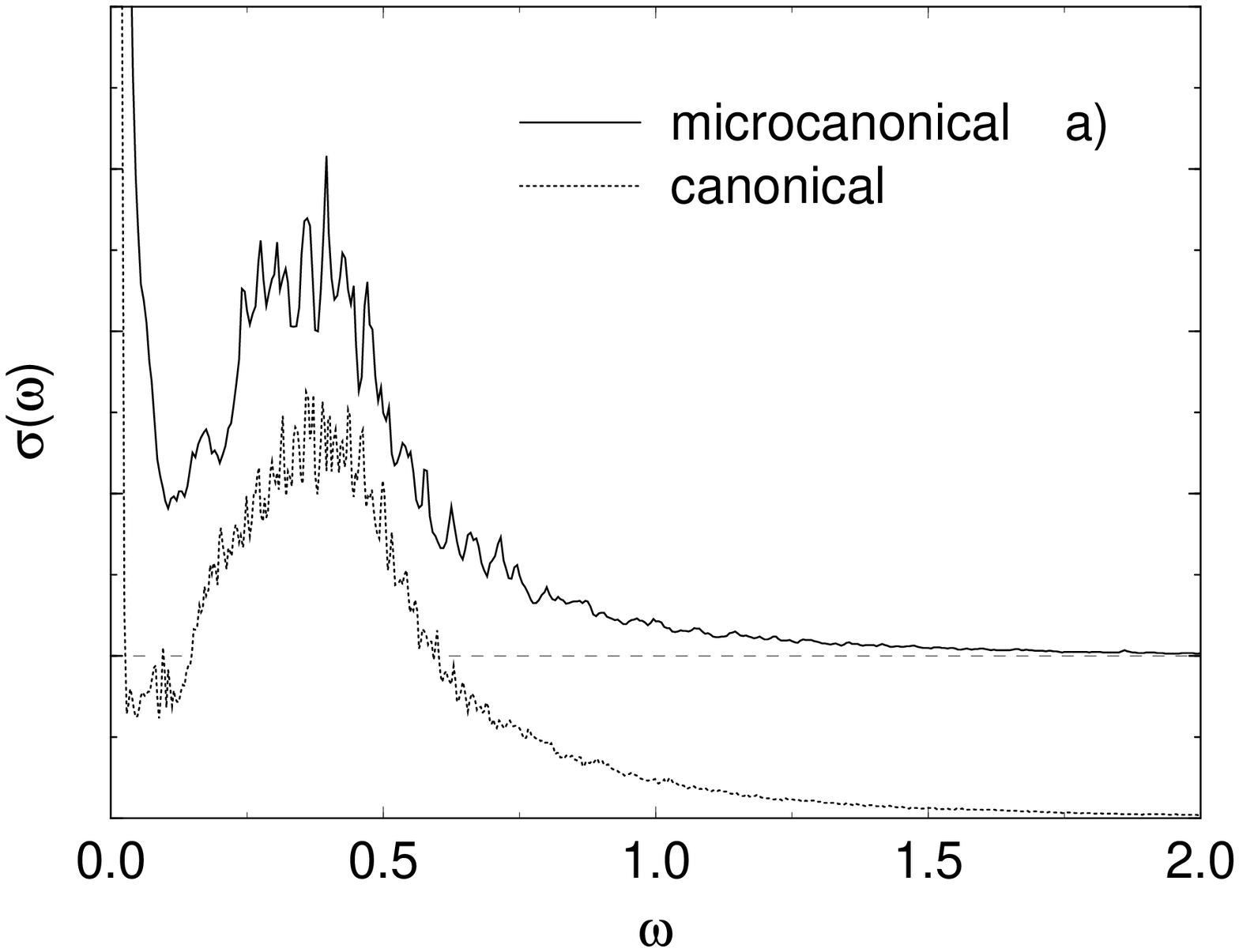}
\includegraphics[width=8.0 cm]{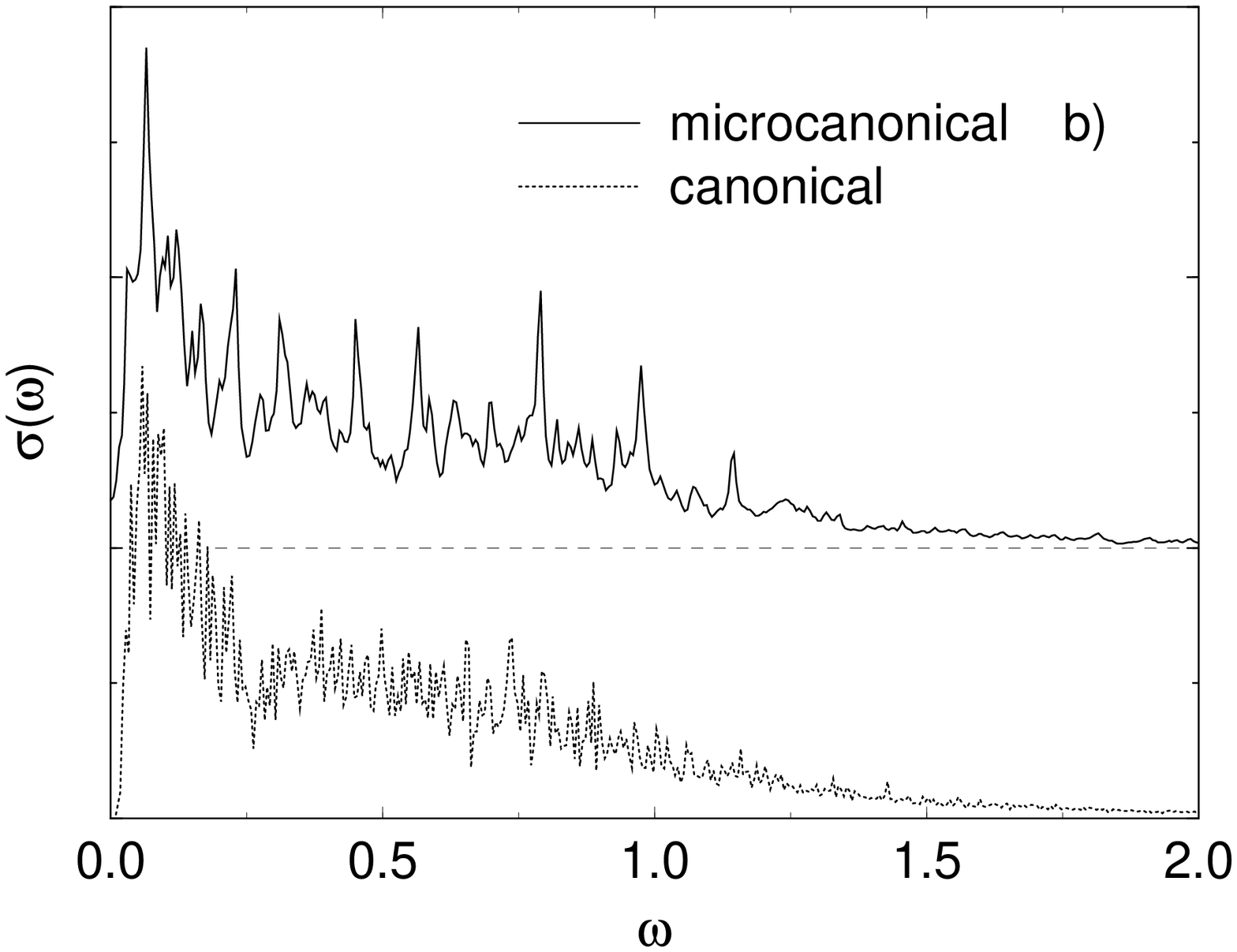}
\caption{
Microcanonical versus Canonical calculations; 
(a) $N=20$, $\Delta =0.5$, $\eta =0.01$,  
(b) $N=18$, $\Delta =1$, $\eta =0.01$.} 
\label{fig3}
\end{figure} 

In Figure \ref{fig3} we present a comparison, extremely encouraging, 
of some microcanonical calculations with the analogous canonical ones.
There is `noise' in all calculations, the origin and magnitude of which we will 
now discuss.
The canonical calculations are essentially a direct
evaluation of expression (\ref{sigma}), where we applied a ``binning" 
procedure on the $\delta-$function weights over an energy scale of about 0.01.
The number of contributing matrix elements are of the order of the dimension 
$\cal D$ of the Hilbert space squared, ${\cal D}^2$, 
e.g. $10^6-10^8$ $\delta-$ functions, 
with no continuity in the weights. The results are not smooth and the
resulting intrinsic fluctuations are heavily smoothed by 
our binning procedure. In the microcanonical calculations we 
employ our scheme, further averaging over translational symmetry $k-$subspaces.
Now, only of $O(\cal D)$ $\delta-$functions 
are essentially contributing, multiplied by the number of states 
involved in the decomposition of the state $\mid \lambda \rangle$ (a few 
thousand depending on the convergence) and the number of $k-$subspaces.
We could average over initial random states, but we find that this has 
only a small smoothing effect, because the underlying poles are at the same 
energies. 
Notice that the observed fluctuations are not associated with any of our 
different resolution processes which are much smaller than the observed 
scale of fluctuations; they are due to the finite size of our system and thus 
to the effective smaller number of matrix 
elements contributing to the construction of the spectra.  
This seemingly new problem associated with our technique turns out
to be dominant for small system sizes; very soon however 
it becomes negligible as  larger systems are achieved, specially 
considering that the dimension of Hilbert space grows exponentially fast 
with the system size $N$. 

\begin{figure}
\includegraphics[width=8.0 cm]{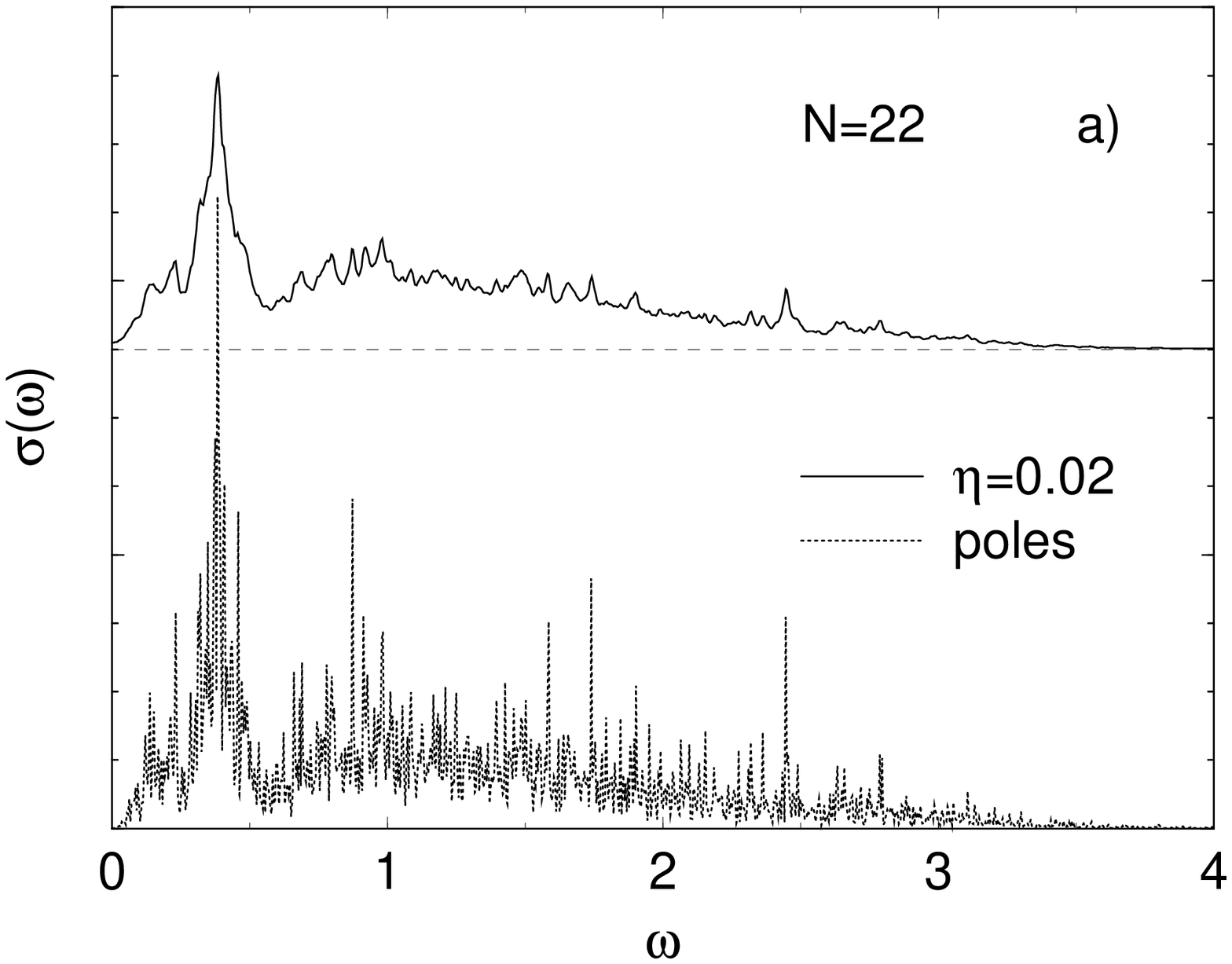}
\includegraphics[width=8.0 cm]{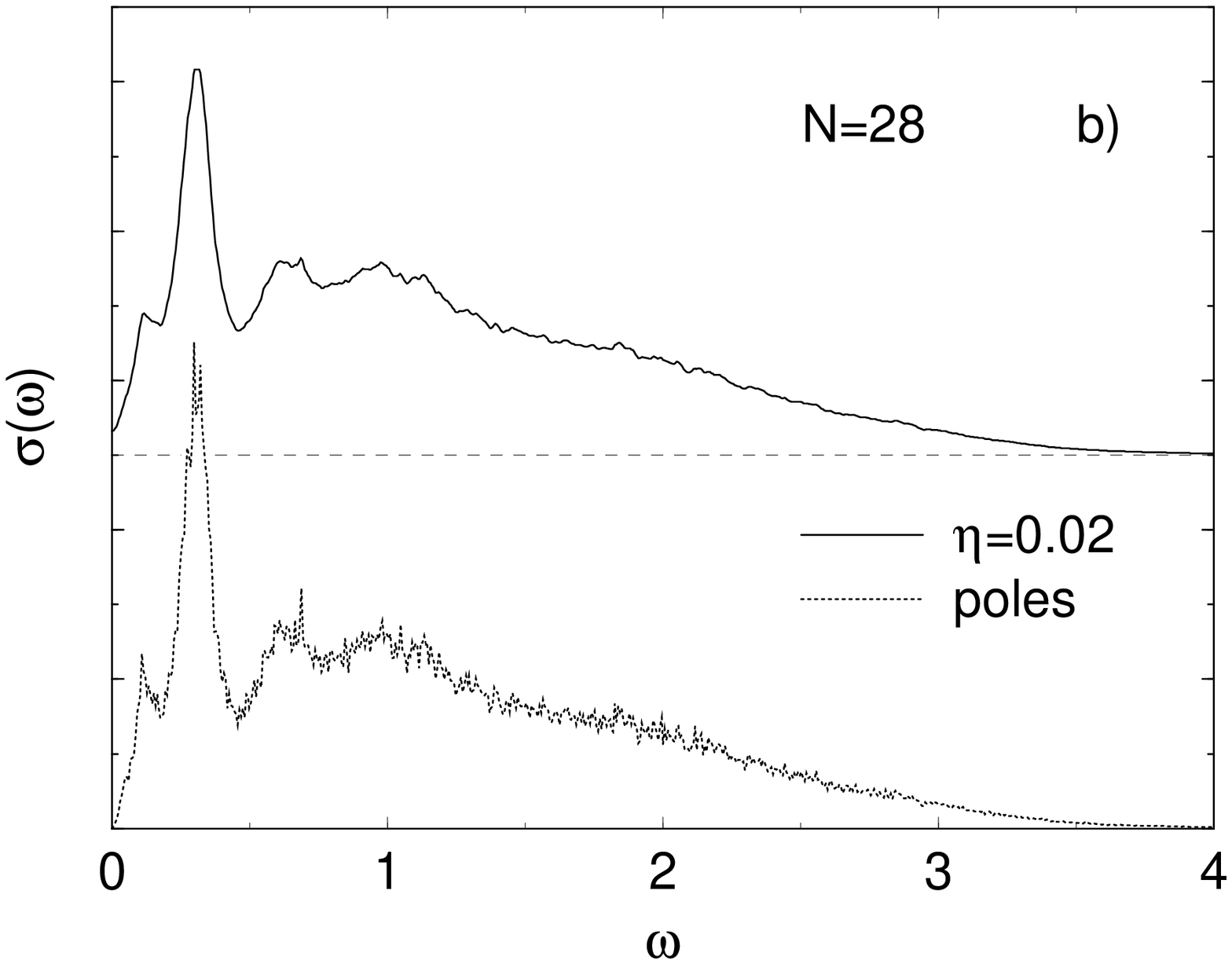}
\caption{
Microcanonical finite-size effects for $\Delta $=2; (a) N=22, (b) N=28.}
\label{fig4}
\end{figure}

In order to
assess these fluctuations and simultaneously the role of our smoothing
parameter $\eta$, we performed some basic calculations involving only a
single $k-$subspace state $\mid \lambda \rangle $. In Figure \ref{fig4} 
we offer a comparison of calculations involving just the poles evaluated 
using the 2nd Lanczos procedure eigenstates
against smoothed versions of the same data but
employing the continued fraction technique. 

The fluctuations clearly decay with system size with the final system 
being surprisingly smooth. The limitations of the
smoothing process are clear, the sharper features are slightly washed out
although the ease of assessing the data makes such a smoothing advisable.
The weights for these microcanonical calculations are truly quite continuous
in comparison to the intrinsic properties of the canonical calculation
which is necessarily ragged.  Obviously for our largest calculations
we are nowhere near converged to the true spectrum which is a possible
explanation for the observed continuity.

We can now fairly safely conclude that our technique is a viable way to
calculate dynamical correlation functions at high temperature for the same
systems accessible by the Lanczos method at $T=0$. By its very
nature, the finite $T$ correlations are much smoother and more regular
to interpret.  Our technique introduces new statistical fluctuations
which make small system sizes ragged but appear to leave large system sizes
essentially unaffected.

\begin{figure}
\includegraphics[width=8.0 cm]{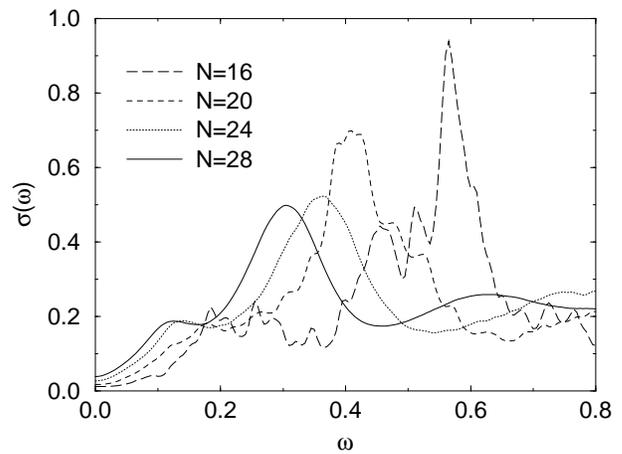}
\caption{
Finite-size scaling for $\Delta =2$}
\label{fig5}
\end{figure}

Although we can now investigate finite temperature dynamic correlations 
using the Lanczos method,
we are still restricted to $N \sim 30$ for a spin-1/2 system. The key to making
useful physical deductions is the procedure of finite-size scaling, the
attempt to deduce the properties of the infinite size system using assumed
properties of the size, $N$, dependence. This method has been widely 
and succesfully applied in the evaluation of ground state energies or gap 
values using data provided by the exact diagonalization, Lanczos or  
Density Matrix Renormalization Group technique.
But to extract information on finite temperature dynamic correlations 
one would need to know the form of the curves before
fitting and scaling could take place mathematically.
As it is clear from Figure \ref{fig5} this might be a challenging task 
considering the statistical fluctuations inherent in the 
spectra \cite{future1}; 
however, from ongoing studies on other systems using 
this method, we find that the behavior of the spectra might greatly depend 
on the model Hamiltonian and correlations under study (e.g. it is far more 
structurless for energy current dynamic correlations). 
Note that the high frequency behavior is generally rather weakly 
size dependent while the low frequency one is 
the most subtle to determine. The last however is the most physically 
interesting as it determines, for instance, the diffusive or ballistic 
behavior of the conductivity.

\begin{figure}
\includegraphics[width=8.0 cm]{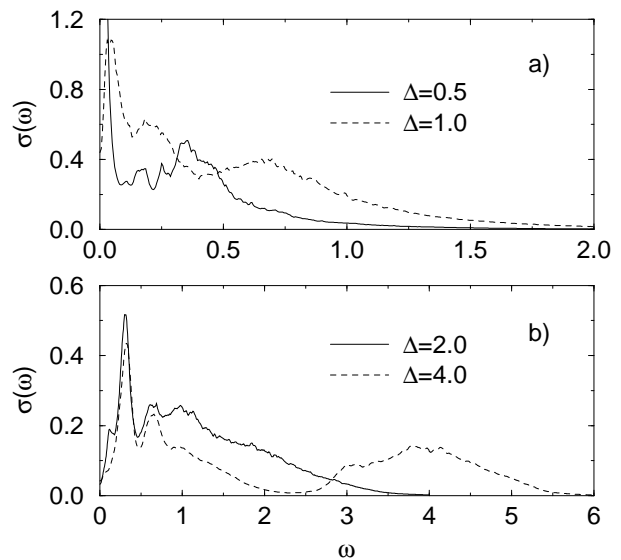}
\caption{
Microcanonical ensemble evaluation of the normalized conductivity 
$\sigma(\omega)$ for $\beta \rightarrow 0,~~~N=28 $;
(a) $\Delta =0.5, 1.0$, (b) $\Delta =2.0, 4.0$}
\label{fig6}
\end{figure}

The basic properties of the $\beta\rightarrow 0$ current-current correlations
are now available and so we provide in Figure \ref{fig6} a few examples 
of the frequency dependence of the conductivity at $\beta \rightarrow 0$ 
as a function of $\Delta$. 

\begin{figure}
\includegraphics[width=8.0 cm]{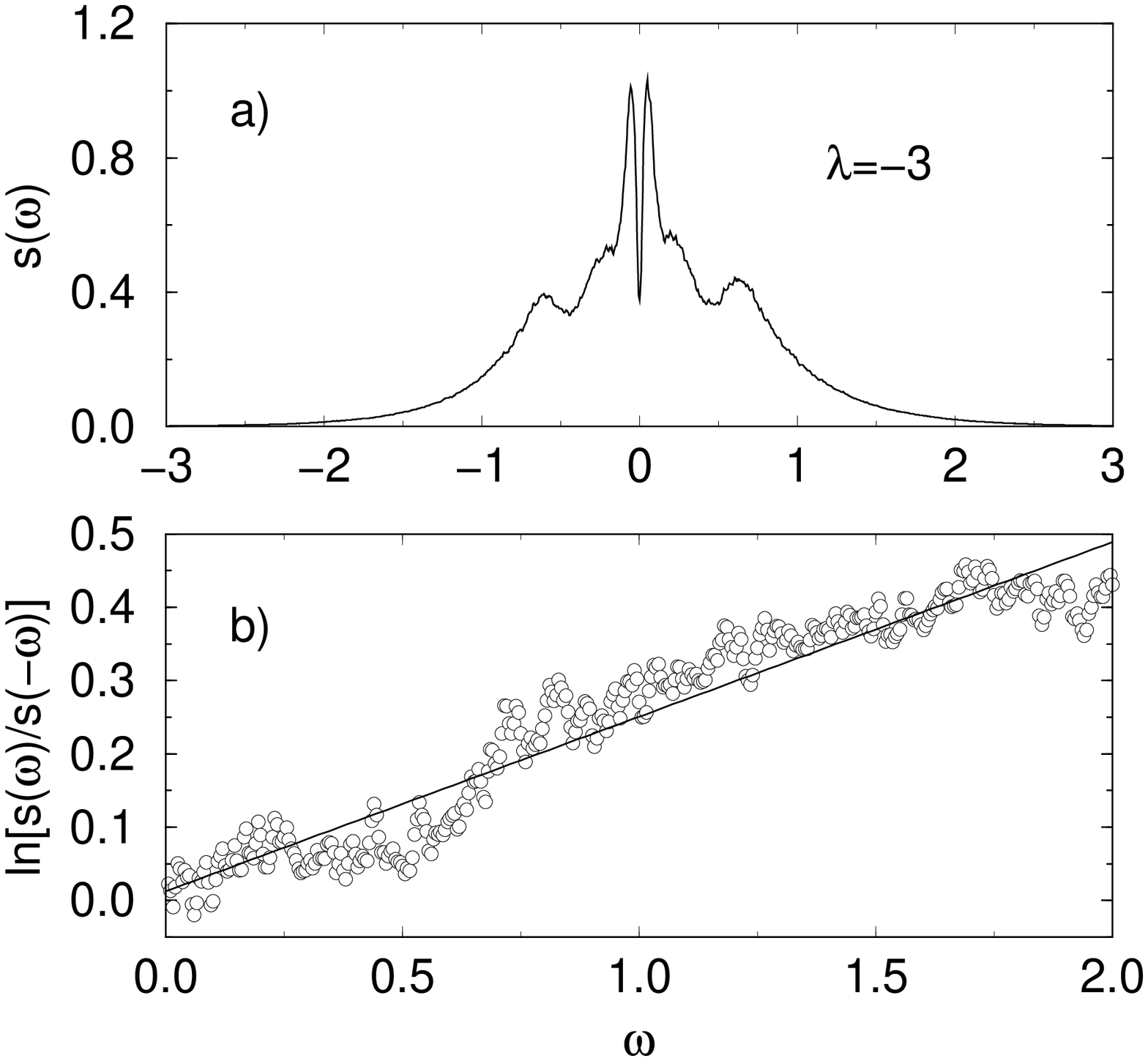}
\includegraphics[width=8.0 cm]{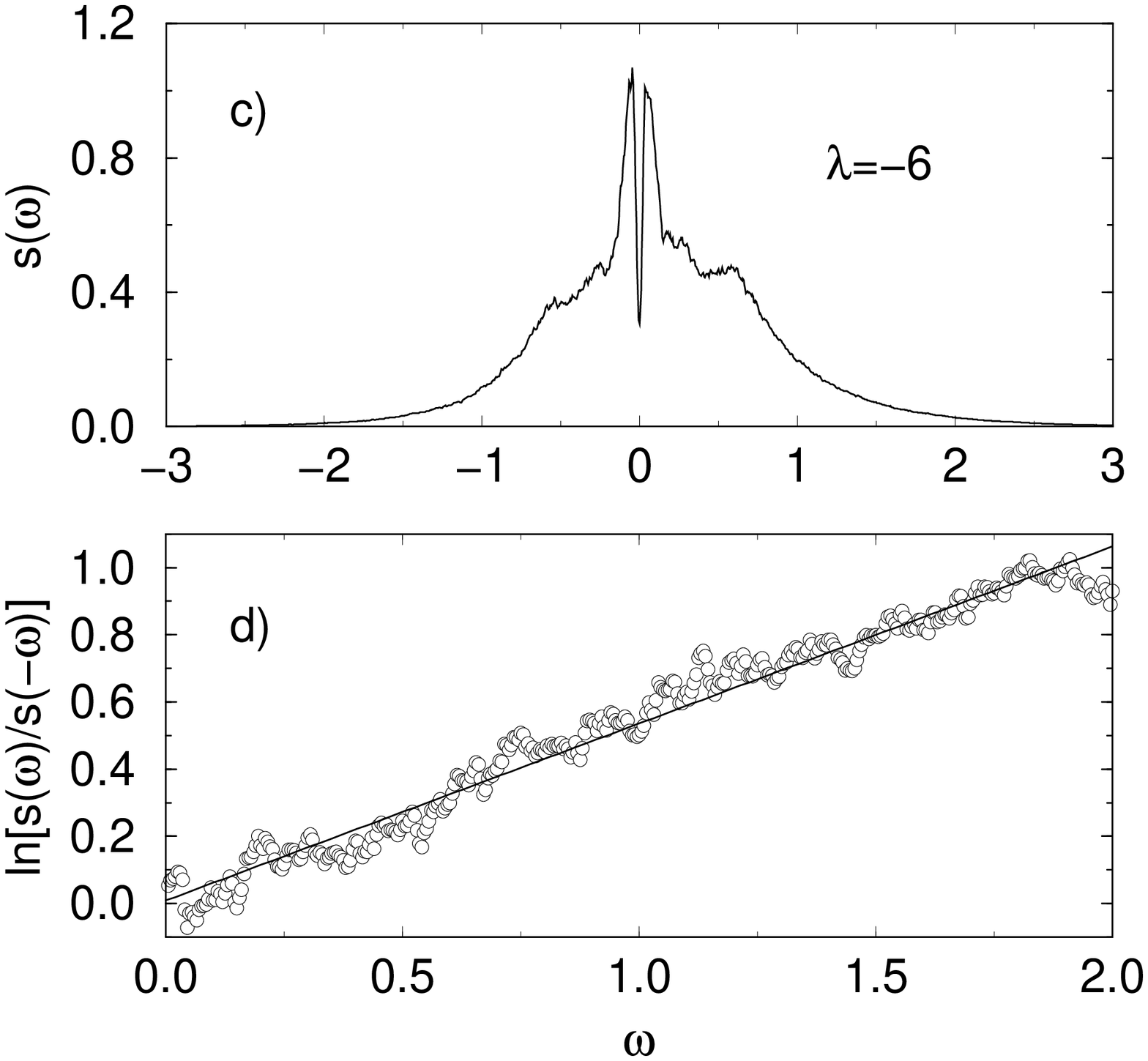}
\caption{
Finite temperature calculations for $N=24$, $\Delta $=1,
$\eta=0.01$; (a) $s(\omega)$, $\lambda=-3$, (b)
Temperature fit $\beta_{micro}\simeq 0.14$,
(c) $s(\omega)$, $\lambda=-6$, (d) Temperature fit $\beta_{micro}\simeq 0.28$.}
\label{fig7}
\end{figure}

Although we have devoted most of our effort to infinite temperature 
($\beta\rightarrow 0$), our technique is valid at essentially any 
temperature (provided that we remain at a dense region of the spectrum). 
Analysing the pure Heisenberg model, we look at a couple of finite 
temperature $k-$averaged calculations in Figure \ref{fig7}.
The temperature has been deduced from a least-squares fit of the quantity,

\begin{equation}
\ln \frac{s(\omega )}{s(-\omega )}\sim \alpha +\beta_{micro} \omega 
\end{equation}

\noindent
to a linear Ansatz, and although the statistical fluctuations are compounded,
an almost vanishing intercept and a clear slope indicate the feasibility of the
strategy. The obtained $\beta_{micro}$ values compare favorably with those 
corresponding to the canonical ensemble in the thermodynamic limit, 
evaluated using $\lambda=< H >_{\beta}$; 
for $\lambda=-3$, $\beta_{micro} \sim 0.14$ vs. $\beta_{canonical}\sim 0.15$, 
for $\lambda=-6$, $\beta_{micro} \sim 0.28$ vs. $\beta_{canonical}\sim 0.3$. 

Although we have compared numerical evaluation of dynamic correlations 
obtained by a canonical and microcanonical method, we have
yet to compare with an exact solution. Recently even non-zero
temperature dynamical correlations have become partially accessible, with a
calculation of the Drude weight for the $0 < \Delta < 1$ Heisenberg model 
at finite temperature \cite{z}. In particular, the 
Drude weight in the $\beta \rightarrow 0$ limit is given 
analytically \cite{kluemperpc} by,

\begin{equation}
D/\beta=\frac{1}{2}\frac{(\pi/\nu -0.5\sin (2\pi/\nu))}{8\pi/\nu},~~~
\Delta=\cos(\pi/\nu).
\label{cjj}
\end{equation}

\noindent
The Drude weight, strictly speaking, is defined as the weight of a zero 
frequency $\delta-$function, eq.(\ref{real}); 
it is a particularity of the Heisenberg model that  
it appears as a narrow 
peak at low frequencies, of the order of the inverse lattice size \cite{nz},  
in contrast to the fermionic ``t-V" version where it is accounted for 
only by the diagonal energy elements ($\omega=0$). 

\begin{figure}
\includegraphics[width=8.0 cm]{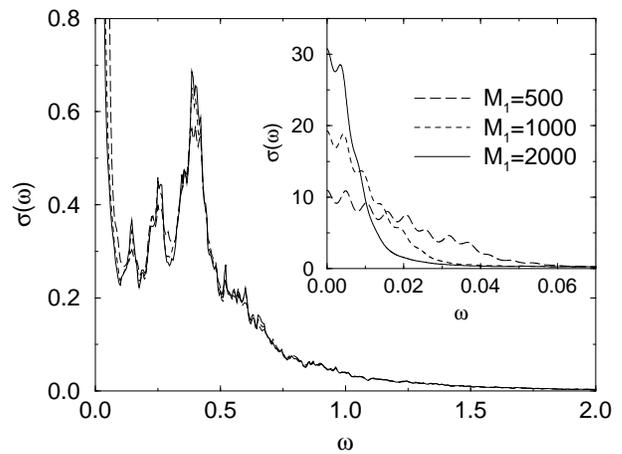}
\caption{
A comparison of three `microcanonical' distributions
$\langle K\rangle =0.002 (M_1=500), 0.0005 (M_1=1000), 0.0012 (M_1=2000)$, 
for $N=26$ and $\Delta=0.5$; inset, low frequency range.}
\label{fig8}
\end{figure}

In extracting the Drude weight by the above described procedure 
we must take into account the 
problem caused by the intrinsic resolution of our calculations, which is of
order $\sigma=\sqrt{\langle K\rangle}$. Although our chosen
resolution of $\sigma \sim 0.01 $ is almost invisible for the smooth 
background, for the Drude weight the resolution is essentially limited by
that of our `microcanonical' distribution, viz $\sigma $.  An example of these
ideas is provided in Figure \ref{fig8},
from which it is clear that the Drude peak is the only contribution for which
the change in resolution is relevant. These calculations involve a single
state and are much improved by $k-$averaging, also the energy window is
so small that the individual poles in the 2nd Lanczos procedure are visible 
and have been smoothed out with an $\eta $=0.005 which adds to the observed 
resolution. In the inset, the scale of the conductivity clearly 
signals a low frequency peak (notice the difference in scale between 
Figure \ref{fig8} and its inset); still in order to extract the Drude weight 
from the smooth background, we must integrate the peak up to at least
as far as it is resolved and that necessitates the inclusion of some of the
background.  We have elected to err on the side of inclusion and tend to
integrate past where the Drude peak appears to become small.

\begin{figure}
\includegraphics[width=8.0 cm]{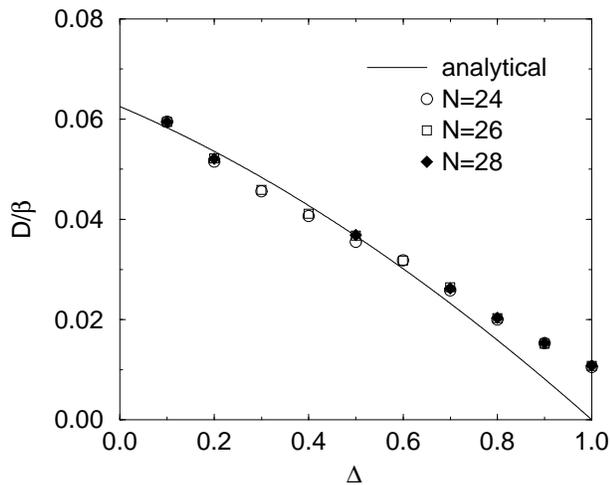}
\caption{
Comparison of $\beta \rightarrow 0$ Drude weight, $D/\beta$;
numerical evaluation (points) vs analytical expression 
eq. (\ref{cjj}) (continuous line).}
\label{fig9}
\end{figure}

In Figure \ref{fig9} we offer a comparison of the analytical and 
numerically extracted Drude weights in the $\beta\rightarrow 0$ limit. 
The quantitative agreement is reasonably satisfactory,  
becoming rather poor near $\Delta \sim 1$ because of 
our technique for extracting the Drude weight; due to the finite
resolution of our calculation we need to sample a finite width around
$\omega =0$.  For the case $\Delta =1$ there is no Drude weight but there does
appear to be a power-law like divergence which we pick up in our finite window
leading to the observed corrupted behaviour.

\section{Discussion}

Our investigation appears to validate the use of the Lanczos algorithm to
analyse finite temperature dynamical properties of strongly correlated
systems; the crucial step is to employ the microcanonical ensemble,
which essentially allows the thermodynamic average to be replaced by an
elementary expectation value.  All the simplicity of the zero temperature
formalism can then be taken over to the finite temperature calculation. The
comparison of canonical with microcanonical procedures 
indicates that the thermodynamic
limit is reached with quite modest system sizes and consequently there appears
to be little systematic error coming from our choice of ensemble. 
There are intrinsic statistical fluctuations in our calculations but these
are severely curtailed by increasing the system size and are an implicit 
difficulty with canonical calculations too. We believe that we
can calculate the high enough temperature dynamical correlations for a finite
system with an excellent tolerance.

The statistical fluctuations in our results require to be
controlled if an error analysis is to be contemplated.  Although we have not
got analytical control, we do have experience at various approaches to
reducing the statistical fluctuations. The crucial point is that,
when taking a statistical average, one should use ``orthogonal" states 
($\mid \lambda \rangle$'s decomposed into different 
sets of eigenstates $\mid n\rangle$). 
Averaging over random starting vectors in
the same subspace is not very effective, even if they are originally
orthogonal, because the resulting distribution involves the same states and
consequently an overlap.  Performing a $k-$ average, 
over translational symmetry subspaces, is an excellent
procedure, since the states are automatically orthogonal and intellectually
one is reverting back towards the real physical statistical average.  Another
possibility is to use several of the eigenstates of the first 
Lanczos procedure; although the orthogonality is guaranteed, there is an 
induced loss in resolution due to the larger $\sigma $'s of the higher 
Lanczos states. A final possibility is to employ the parameter $\lambda$, 
where the average over different $\lambda$'s must be limited within a 
window that corresponds to the energy fluctuations at the studied 
temperature in the given size system. Providing
that the $\lambda$'s are further apart than the chosen $\sigma$, the
orthogonality is essentially guaranteed. 

Although we believe we have access to the temperature behaviour of 
finite-size systems, this does not give immediate access to the dynamics 
in the thermodynamic limit because finite-size scaling must be performed; 
Figure 6 exhibits clear peaks of unknown form,
plausible `cusps' and regions where the correlations vanish.  Unless we can
guess or deduce the form of these structures, finite-size scaling appears
problematic. We should note however from our experience, that not 
all models and dynamic correlations exhibit so involved spectra;  
in forthcoming works we will present analysis of charge/spin/energy current 
correlations for other (non-) integrable systems of current interest  
(higher spin, ladder models) where the obtained spectra are far more 
structrurless.
Finally, besides the finite frequency behavior, our method allows the 
reliable study of scalar quantities as the Drude weight. 

%\appendix

\begin{acknowledgments}
Part of this work was done during visits of (P.P.) and (M.L.) at IRRMA as
academic guests of EPFL.
J.K. and X.Z acknowledge support by the Swiss National Foundation, 
the University of Fribourg and the University of Neuch\^atel.
\end{acknowledgments}


\begin{references}
\bibitem{future1} The low frequency behavior of the spin-1/2 
Heisenberg model, in particular the finite-size dependence  
as extracted from such a microcanonical study, will be adressed 
in a forthcoming publication.
\end{references}
\end{document}